\documentclass[twocolumn, trackchanges]{aastex631}

\usepackage{graphicx}
\usepackage{float}
\usepackage[tight,TABTOPCAP]{subfigure}
\usepackage{hyperref}
\usepackage{xurl}
\usepackage{enumitem}
\usepackage{amsmath}
\usepackage{amsfonts}
\usepackage{amssymb}
\usepackage{afterpage}
\usepackage{savesym}
\savesymbol{tablenum}
\usepackage{siunitx}
\restoresymbol{SIX}{tablenum}
\usepackage{amsmath}
\usepackage{color}
\usepackage{xcolor}
\usepackage[toc,page]{appendix}

\begin{document}
\title{\LARGE \textbf{Physics-Informed Machine Learning for Modeling \newline
                      Turbulence in Supernovae}}
\author{Platon I. Karpov} 
\affiliation{Department of Astronomy $\&$ Astrophysics, University of California, Santa Cruz, CA 95064, USA}
\affiliation{Los Alamos National Laboratory, Los Alamos, NM 87545, USA}
\author{Chengkun Huang} 
\affiliation{Los Alamos National Laboratory, Los Alamos, NM 87545, USA}
\author{Iskandar Sitdikov}
\affiliation{Provectus IT Inc., Palo Alto, CA 94301, USA}
\author{Chris L. Fryer}
\affiliation{Los Alamos National Laboratory, Los Alamos, NM 87545, USA}
\author{Stan Woosley}
\affiliation{Department of Astronomy $\&$ Astrophysics, University of California, Santa Cruz, CA 95064, USA}
\author{Ghanshyam Pilania}
\affiliation{Los Alamos National Laboratory, Los Alamos, NM 87545, USA}
\begin{abstract}
Turbulence plays an important role in astrophysical phenomena, including core-collapse supernovae (CCSN), but current simulations must rely on subgrid models since direct numerical simulation (DNS) is too expensive. Unfortunately, existing subgrid models are not sufficiently accurate. Recently, Machine Learning (ML) has shown an impressive predictive capability for calculating turbulence closure. We have developed a physics-informed convolutional neural network (CNN) to preserve the realizability condition of Reynolds stress that is necessary for accurate turbulent pressure prediction. The applicability of the ML subgrid model is tested here for magnetohydrodynamic (MHD) turbulence in both the stationary and dynamic regimes. Our future goal is to utilize this ML methodology \href{https://github.com/pikarpov-LANL/Sapsan/wiki/Estimators\#physics-informed-cnn-for-turbulence-modeling}{(available on GitHub)} in the CCSN framework to investigate the effects of accurately-modeled turbulence on the explosion of these stars.
\end{abstract}

\section{Introduction}

Turbulence plays a key role in many astrophysical phenomena \citep{Schekochihin2007, brandenburg2013, beresnyak2019review}. A prominent example being a core-collapse supernova (CCSN): the
bright, energetic, dynamic explosion of a highly evolved massive star
of at least 8 times the mass of the sun. At the end of its life, the
iron core of such a massive star can no longer generate energy by
fusion reactions and yet is subject to enormous energy losses in the
form of neutrinos. As the core of about 1.5 solar masses contracts and
heats up, looking for a new source of energy generation, additional
instabilities instead accelerate the collapse until it is in almost free fall. These instabilities include electron capture and photodisintegration - heavy nuclei splitting into lighter elements due to high-energy photon absorption. The collapse
continues until the density of the inner core exceeds that of the
atomic nucleus ($\sim \num{2e14}\:g\:cm^{-3}$) and then abruptly
halts due to the repulsive component of the nuclear force. The outer
part of the core rains down on the nearly stationary inner core and
bounces, creating a powerful outward-bound shock wave. It was once
thought that this ``prompt shock'' might propagate through the entire
star, exploding it as a supernova \citep{baron1987}. Now we know it does not
happen. The shock stalls in the face of prodigious losses to neutrinos
and photodisintegration and becomes an accretion shock outside the
edge of the original iron core. All positive radial velocity is gone
from the star. The evolution slows, now taking 100's of milliseconds
instead of milliseconds. At this stage, the core is a hot proto-neutron star,
radiating a prodigious flux of neutrinos, surrounded by an accretion
shock through which the rest of the star is falling. The success or
failure of the explosion then depends on the efficiency of neutrinos
in depositing some fraction of their energy outside the proto-neutron
star (outside the neutrinosphere and inside the accretion shock), and the distribution of pressure
that energy deposition creates. A failed explosion will lead to a
black hole and no supernova. \citep{woosley2005ccsn, burrows2021}

Over the past three decades, the community has focused on the importance of turbulence and convection in improving the efficiency with which energy released in the collapse of the core is converted into explosion energy~\citep{1994ApJ...435..339H,2003ApJ...584..971B,2007ApJ...659.1438F,2015ApJ...808L..42M,2018SSRv..214...33B}. Most of these studies focused on the large-scale convective motions that transport both matter and energy. If the pressure in this convective region, including turbulence, becomes large, the accretion shock will be pushed outwards, ultimately achieving positive velocity and exploding the star. A recent study attributes up to $\sim30\%$ of the gas pressure to turbulence to aid the CCSN explosion \citep{nagakura19}. Turbulence in this region has three origins: the primordial
turbulence present because the star was convective in these zones
before it collapsed; the turbulence generated by the convective overturn
driven by neutrino energy deposition beneath the shock; and, if the
star is rotating, by magnetic instabilities in the differentially
rotating layers (especially the ``magneto-rotational instability'',
MRI). Multi-dimensional solutions exist to the CCSN problem both with
and without rotation and magnetic fields. Some explode; some do not,
and this has been a problem for at least the past 60 years \citep{Colgate1966}. A major difficulty is a physically correct
description of the turbulence and its effective pressure in a
multi-dimensional code that is unable to resolve the relevant length
scales.

Here we focus on magnetically generated turbulence. This introduces
additional variables and uncertainties not contained in the non-MHD
case, but has the merit of using conditions that are locally generated and
the existence of a high-resolution training set \citep{moesta2015} (detailed in Section \ref{sec:datasets}).
The framework that we derive can be used for both MHD and field-free
turbulence, and it is our intention to return to the non-MHD case in subsequent work. In this case, the MRI occurs in a setting of magnetized, differentially rotating fluid layers, i.e., stellar shells.
The instability exponentially amplifies primordial perturbations
developing turbulence \citep{obergaulinger2009mri}.

A flow can be considered turbulent if the Reynolds number ($Re$) is of order $\sim10^3$, which corresponds to what we expect to see in CCSN \citep{fryer2007}. For DNS, the per-axis 3D resolution scales as $N \sim 2Re^{3/2}$ \citep{jimenez2003}, leading to 3D DNS of CCSN requiring a grid size of $10^{5}$ in each direction, which is extremely expensive (if possible) to achieve with today's HPC resources. Together with a vast spatial scale range needed to be resolved in CCSN ($200\:km$ inner convective region and out to $10^9\:km$ outer shell) and the complexity of physical processes ongoing in CCSN, DNS calculations are out of reach. Given the computational challenges, subgrid turbulence is often modeled using the following techniques (in astrophysics, these schemes are primarily used in 1-dimensional simulations):

\begin{itemize}
    \item \textbf{Mixing Length Theory (MLT)} - modeling turbulent eddies that transfer their momentum over some \textit{mixing length} via eddy viscosity \citep{spiegel1963mlt}; akin to molecular motion. MLT is used to study turbulence driven by convection, thus applicable to stellar convection (including supernovae simulations \citep{Couch_2020}) in 1D simulations. MLT performs well for small mixing length scales, while turbulence in CCSN evolves over a wide range of scales, which is deemed problematic for MLT's accuracy \citep{mlt_limits}.
    \item \textbf{Reynolds-Averaged Navier Stokes (RANS)} - time (and ensemble) averaged treatment of turbulence equations of motion. RANS is typically used for relatively low $Re$, e.g., stellar evolution and consequently supernovae progenitors \citep{Arnett_2015} and requires a turbulent closure model.
    \item \textbf{Large Eddy Simulation (LES)} - space averaged turbulence treatment. Similar to RANS, a subgrid model substitutes the turbulence effects absent from small spatial scales. For \textit{closure}, it is common to use Dynamic Smagorinsky \citep{lilly1966} or gradient-type subgrid models \citep{schmidt2015les, miesch2015mhd} in LES simulations.
    \textbf{Implicit LES (ILES)} - similar to LES, but the small scales are assumed to be approximated by numerical artifacts (e.g., numerical diffusion and viscosity). ILES is typically employed by large, global 3D simulations of CCSN and other astrophysical events \citep{radice2015, radice2018}.
\end{itemize}

\begin{figure}
    \centering
    \includegraphics[width=\linewidth]{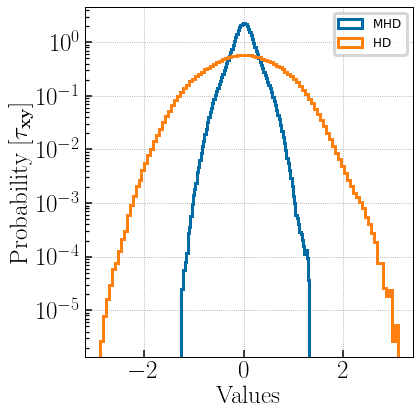}
    \caption{An example of the difference magnetic field can bring to the turbulence stress distribution. The data has been taken from JHTDB MHD and HD datasets \citep{jhtdb2008}.}
    \label{fig:hd_mhd}
\end{figure}

Even though these techniques have achieved some level of success when predicting HD turbulence, MHD poses a new set of challenges. With magnetic fields present, the magnetic/kinetic energy is primarily transferred at small scales through the dynamo process \citep{beresnyak2014dynamo}. The non-linear behavior of MHD further exacerbates simulation challenges leaving many open questions on the nature of turbulence despite decades of focused studies \citep{beresnyak2019review}. As an example of changes due to an introduction of a magnetic field within a simulation, we present a comparison of normalized $Re$ stress tensor component distribution between HD and MHD simulations given comparable initial conditions from Johns Hopkins Turbulence Database (JHTDB) in Figure \ref{fig:hd_mhd}.

To tackle the challenges of MHD turbulence in an astrophysical setting, we turned to Machine Learning (ML). In HD simulations, ML has shown promising results in the fields of applied mathematics, engineering, and industry as a whole. Throughout the last few decades, there have been significant technological and algorithmic advancements that have started a new era for the study of turbulence through the means of \textit{Big Data}. For our context, by \textit{Big Data} we mean the abundance of turbulence DNS data that resulted from the Moore's Law evolution of high-performance computing (HPC). While $Re$ of those simulations is not to the natural level of CCSN, it is still a significant improvement upon the resolution of the current CCSN models. Furthermore, DNS data in non-astronomy fields has been used as ground-truth, along with experimental data, for turbulence subgrid model optimization in large-scale 3D computational fluid dynamic (CFD) simulations. \textit{Big Data} presents the opportunities for ML to help augment and improve turbulence modeling. As shown by the industry (i.g., face-recognition, self-driving cars), ML is highly flexible, and it has already shown its potential for turbulence modeling, both direct prediction of turbulent fluxes and analytical (e.g., RANS-based) model optimization \citep{king2016, rosofsky2020mlastro, wu2018, zhang2018}.

ML has been applied to quasi-stationary 2D ideal MHD turbulence in an astrophysical setting with promising results as compared to the conventional analytic gradient subgrid turbulence model \citep{rosofsky2020mlastro}. In this paper, we developed an ML model for 3D MHD turbulence for astrophysical simulations in reduced dimension, as well as performed a time-dependent prediction analysis. Considering the popularity and success of Convolutional Neural Networks (CNNs) in the industry \citep{AlexNet, ResNet} for spatial pattern recognition, we sought to relate large-scale eddie structure to small-scale turbulence distribution. Thus, we based our approach on CNNs to develop a physics-informed machine learning (PIML) turbulence closure model, described in Section \ref{sec:piml}. It was applied for two regimes to test generalizability: statistically-stationary homogeneous isotropic MHD turbulence from a general-purpose dataset and dynamic MHD turbulence from a CCSN simulation. The former can be found in the ISM, studying the stellar formation, and the latter is directly applicable to high-energy events, such as CCSN. The model predicts Reynolds stress tensor ($\tau$), with turbulent pressure ($P_{turb}$) defined as:
\begin{gather}
    P_{turb} = tr(\tau)
    \label{eq:pturb}
\end{gather}
\noindent where $tr(x)$ is a trace of $x$. Note that we neglect the $1/3$ coefficient in our definition of $P_{turb}$, which has no effect on the ML prediction results. We will primarily focus on analyzing statistical distribution, i.e., probability density function (PDFs), of time-dependent turbulence. In the case of stationary turbulence, we will be testing the stability of the ML model, ensuring prediction would remain in the physical domain. For the dynamic case, we will check the model's ability to make future predictions within the limits of the available \textit{ground truth} data, i.e., DNS data we assume to accurately represent the physical state of the system.

In Section \ref{sec:formalism}, we will cover filtering, decomposition, and the result MHD formalism. Section \ref{sec:subgrid} introduces the analytical subgrid turbulence model we will be using for comparison, our ML pipeline, specifics of the ML \& PIML models used to treat various $\tau_{ij}$ components, and the datasets used for training and testing them. In Section \ref{sec:results}, we provide the analysis of stationary and dynamic results, with conclusion following in Section \ref{sec:conclusion}. Lastly, the Appendices include further details on the ML model developed and its training process.

\section{Formalism}
\label{sec:formalism}
We begin by presenting the mathematical basis of our work, covering the fundamentals of MHD LES, including its unfiltered and filtered forms.

\subsection{Filtering}
\label{sec:filtering}
A filtering operation is defined as an infinitely-resolved, i.e., continuous, variable that is decomposed into \textit{average} and \textit{fluctuating} parts:
\begin{gather}
    u = \bar{u}+u'
\end{gather}
where $\bar{u}$ (LES-simulated quantity) is the ensemble average of $u$ (DNS quantity), and $u'$ are the fluctuations. By cutting out \textit{fluctuations} of a specific size, what is left can be thought of as a filtered quantity. Then, $\bar{u}$ is defined by applying a filtering convolution kernel $G$:
\begin{gather}
    \bar{u} = G\ast u
\end{gather}
In the context of LES, the simulation resolution is defined as a spatial filter of size $\Delta$ applied to a continuous variable. What we have done in this study is typical for the LES community: taking a high-resolution DNS data and applying a filter of size $\Delta_f$, where $\Delta_f > \Delta$, to decrease (``blur'') the fidelity of the data to mimic a low-resolution simulation. 

We applied a Gaussian filter to all of the data, with a 1D Gaussian kernel as $G$:
\begin{gather}
    G(x) = \frac{1}{\sqrt{2\pi\sigma^2}}e^{-\frac{x^2}{2\sigma^2}}
    \label{eq:gauss}    
\end{gather}
where $\sigma$ is the standard deviation of the Gaussian, controlling the amount of ``blur'', and $x$ is the data. The filter can be applied in 3D via the product of 1D Gaussian functions, covering each direction.

\subsection{MHD equations - Unfiltered}

In order to bridge the gap between filtered and unfiltered values, i.e., a stress tensor that we will model, let's establish the basis MHD Navier-Stokes equations.
\begin{eqnarray}
    &\partial_t\rho+\partial_i[\rho u^i] = 0 \label{eq:mhd1} \\
    &\partial_t(\rho u^j)+\partial_i\bigg[\rho u^i u^j -B^iB^j+\delta^{ij}\bigg(\rho+\frac{B^2}{2}\bigg)\bigg] = 0 \label{eq:mhd2} \nonumber \\
    &\partial_tB^j+\partial_i[u^iB^j-u^iB^i] = 0 \label{eq:mhd3} \nonumber \\
    &\partial_iE+\partial_i[(E+p+B^2)u^i-(u_jB^j)B^i] = 0 \label{eq:mhd4} \nonumber
    &\label{eq:mhd}
\end{eqnarray}
\noindent where $\rho$ is density, $p$ is pressure, $e$ is internal energy density, $u^i$ is velocity, $B^i$ is a magnetic field, $\delta^{ij}$ is the Kronecker delta, and the indices are spatial axes. In addition, total energy density $E$ is defined as follows:
\begin{gather}
    E = e+\frac{\rho v^2}{2}+\frac{B^2}{2}
\end{gather}

\subsection{MHD equations - Filtered}
\label{sec:mhd_filt}
For the filtered equations, we need to apply Eq. \ref{eq:gauss} to Eqs. \ref{eq:mhd1}. As a result we get \citep{vigano2019}:
\begin{eqnarray}
    &\hspace*{-0.3cm}\partial_t \bar{\rho}+\partial_i[\bar{\rho}\tilde{u}^i]=0 \\
    &\hspace*{-0cm}\partial_t(\bar{\rho}\tilde{u}^j)+\partial_i\bigg[\bar{\rho}\tilde{u}^i\tilde{u}^j-\bar{B}^i\bar{B}^j+\delta^{ij}\bigg(\tilde{p}+\frac{\bar{B}^2}{2}\bigg)\bigg] = -\partial_i \tau_{mom}^{ij} \nonumber \\
    &\hspace*{-0.3cm}\partial_t\bar{B}^j+\partial_i[\tilde{u}^i\bar{B}^i-\tilde{u^j}\bar{B^i}] = -\partial_i\tau_{ind}^{ij} \nonumber \\
    &\hspace*{-0cm}\partial_t\bar{E}+\partial_i[(\bar{E}+\tilde{p}+\bar{B}^2)\tilde{u}^i-(\tilde{u}_j \bar{B}^j)\bar{B}^i] = -\partial_i\tau_{eng}^i+\Sigma_{eng} \nonumber
\end{eqnarray}
\noindent where $\tau$ stands for various sub-grid scale (SGS) tensors. Defined as follows: [include full definitions]
\begin{eqnarray}
    &\hspace*{-0.3cm}\tau_{mom}^{ij} = \bar{\rho}\tau_{kin}^{ij}-\tau_{mag}^{ij}+\delta^{ij}\bigg(\frac{1}{2}\delta_{kl}\tau_{mag}^{kl}+(\bar{p}-\tilde{p})\bigg) \\
    &\hspace*{-0.3cm}\tau_{eng}^i = \tau_{enth}^i+\tau_{mom}^{ij}\tilde{u}_j+\tau_{ind}^{ij}\bar{B}_j
\end{eqnarray}
\noindent lastly, the SGS tensors represent:
\begin{eqnarray*}
    \tau_{kin} &:& \text{turbulent motion} \\
    \tau_{mag} &:& \text{B contribution to the motion} \\
    \tau_{ind} &:& \text{turbulent amplification of $B$} \\
    \tau_{enth} &:& \text{turbulent contribution to energy transfer}
\end{eqnarray*}
In this paper, we focused on modeling $\tau_{kin}$, i.e., Reynolds stress:
\begin{gather}
    \tau_{kin}^{ij}=\widetilde{u^iu^j}-\tilde{u}^i\tilde{u}^j
    \label{eq:velocity_stress}    
\end{gather}
Throughout the rest of the paper, $\tau_{kin}^{ij}$ will be referred to as $\tau_{ij}$ to simplify notation.

\section{Subgrid Modeling}
\label{sec:subgrid}
We will be comparing our ML results with a conventional turbulence subgrid model used widely in astronomy - the gradient model \citep{liu1994}.

\subsection{Gradient model}

\label{sec:gradient}

The gradient model is defined by the Taylor series expansion of the filtering operation. The tensor has the form of:
\begin{gather}
    \tau_{ij} = \frac{\tilde{\Delta}^2}{12}\delta_k\tilde{u}_i\delta_k\tilde{u}_j
\end{gather}
where $\tilde{\Delta}$ is the filter size, and $u$ is velocity.

\begin{figure*}
  \centering
  \subfigure[Off-diagonal terms (3D CNN)]{\includegraphics[width=0.385\linewidth]{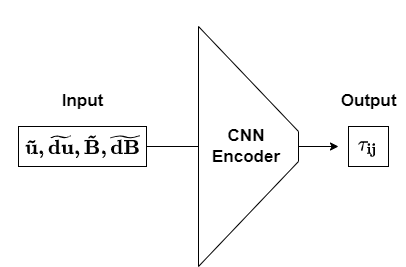}}
  \quad
  \subfigure[Diagonal terms (PIML)]{\includegraphics[width=0.5\linewidth]{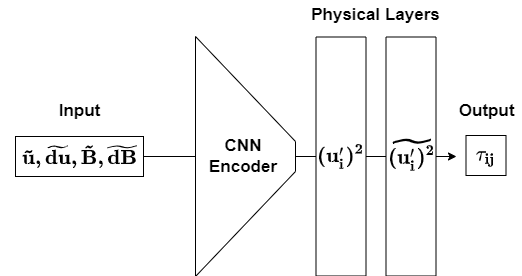}}
  \label{fig:piml_graph}
  \caption{Model schematics to calculate the Reynolds stress ($\tau_{ij}$) components.}
  \label{fig:ml_graphs}
\end{figure*}

\subsection{Machine Learning Pipeline}

In its essence, machine learning (ML) is a sophisticated fitting routine of a multi-dimensional dataset against a target feature. However, unlike it, ML does not require a theoretical understanding of the underlying statistical form of the data. Thus, the exact relationship of a feature to the target does not need to be defined in contrast to conventional fitting routines. ML methods are capable of \textit{learning} the data structure solely from the input data with model tuning based on a validation dataset \citep{bishop2006pattern, lecun2015deep}. This opens up a possibility to learn new links between the input variable, potentially leading to new physics and functional forms \citep{carleo2019}. While we will not delve deeper into the latter topic in this paper, we will discuss how to use physics to inform and then further analyze the model training.

Lastly, ML models can learn iteratively, hence improving themselves as new data becomes available. That signals the potential to achieve accurate interpolation/classification and, more importantly, extrapolation results \citep{carleo2019}. This applies to both spatial and temporal data. 

Currently, there are ML models that are based on convolutional neural networks. They are used as generalizable solutions and are standard in the industry (e.g., AlexNet \citep{AlexNet}, ResNet \citep{ResNet}). However, those models are not optimized to solve problems in physical sciences, including astrophysics. Considering the lack of standardized packages for ML in astrophysics, we built our own pipeline, {\sf Sapsan} \citep{Karpov2021}. Here is a high-level procedure overview:

\begin{enumerate}
    \item \textbf{Data}: choose a relevant high-fidelity dataset. In our case, the data comes from the DNS simulations that we consider to be ground truth.
    \item \textbf{Data Augmentation}: filter and augment the data to mimic an LES simulation, in which the ML pipeline would be used. For example, the turbulent features in the CCSN LES simulations are severely under-resolved. Hence filtering applied to high-resolution DNS simulation data need to account for that adequately.
    \item \textbf{Data Splitting}: split the data into training, validation, and testing portions.
    \item \textbf{Optimization and Training}: optimizing hyperparameters of the ML model via cross-validation and testing against the unseen data. In this context, \textit{unseen} data is defined by the data not used in the training or validation of the ML model. This procedure strives to achieve the best efficiency and accuracy of the model.
    \item \textbf{Validation, Testing, and Analysis}: test the trained ML model to confirm the predictions to be representative of the relevant physics, as well as an estimate of efficiency and uncertainty of the ML scheme.
\end{enumerate}

Next, we will discuss how we adopted and augmented conventional ML methods for CCSN turbulence prediction, enforcing physical principles to aid our studies.

\subsection{Machine Learning Models}
\label{sec:ml}

We used two ML models to calculate all 9 components of the Reynolds stress $\tau_{ij}$: a conventional CNN encoder for off-diagonal terms and a custom, physics-informed CNN encoder for diagonal terms. Schematics of the models are shown in Fig.\ref{fig:ml_graphs}a and Fig.\ref{fig:ml_graphs}b with discussion of each in Sections \ref{sec:3dcnn} \& \ref{sec:piml} respectively. Both models have been trained on a dual-GPU system, equipped with NVIDIA Quadro RTX 5000 cards.

\subsubsection{Off-Diagonal Terms (3D CNN)}
\label{sec:3dcnn}
The idea behind a neural network is illustrated in Fig.~\ref{fig:ml_graphs}a as a pipeline schematic. We first need to put in the data, which is represented by the input layer. Then the data will need to be manipulated in some way, as represented by the hidden layer(s), e.g., \textit{CNN Encoder}. At the end, there is an output layer predicting our target quantity.

We based our model for off-diagonal tensor components on the 3D Convolutional Neural Network (3D CNN) with some modifications while keeping it conventional. In a CNN, a \textit{convolution} layer is applied as a hidden layer. In that case, a given kernel is used to parse through the dataset to identify the spatial patterns needed for the given problem. The kernel has the form:
\begin{multline}
    out(N_i, C_{out}) = \\
    bias(C_{out}+\sum_{k=0}^{C_{in}-1}weight(C_{out,k})input(N_i,k))
\end{multline}
where $N$ is the number of features, $i$ is the feature index, and $C$ is the number of channels. The input data size is defined by $[N,C_{in},D,H,W]$, and the output is $[N,C_{out},D_{out},H_{out},W_{out}]$, where [$D$, $H$, $W$] are the [depth, height, width], i.e., [x,y,z]. The notation is in agreement with {\sf PyTorch} documentation$^1$. The reason behind choosing a CNN as our core ML algorithm was the goal to relate spatial structure of the turbulent eddies to the small scale structure.

We utilized {\sf PyTorch} build-in modules with slight modifications for our workflow, with the following parameters:
\begin{itemize}
    \item \textbf{Model}: a classical approach for CNN network architectures, where convolution and pooling layers are stacked up consequently followed by fully connected dense layers as it is shown in Fig.~\ref{fig:3dcnn_graph}, based around 3D CNN\footnote{\href{https://pytorch.org/docs/stable/generated/torch.nn.Conv3d.html}{pytorch.org/docs/stable/generated/torch.nn.Conv3d.html}}.
    \item \textbf{Optimizer}: Adam Optimizer\footnote{\href{https://pytorch.org/docs/stable/generated/torch.optim.Adam.html}{pytorch.org/docs/stable/generated/torch.optim.Adam.html}} - extension of the stochastic gradient descent; it was picked due to the good performance on sparse gradients.
    \item \textbf{Activation function}: LogSigmoig\footnote{\href{https://pytorch.org/docs/stable/generated/torch.nn.LogSigmoid.html}{pytorch.org/docs/stable/generated/torch.nn.LogSigmoid.html}} - a non-linear activation function to select values to pass from layer to layer. The function is defined by $log(\frac{1}{1+exp(-x)})$.
    \item \textbf{Loss function}: Custom SmoothL1Loss\footnote{\href{https://pytorch.org/docs/stable/generated/torch.nn.SmoothL1Loss.html}{pytorch.org/docs/stable/generated/torch.nn.SmoothL1Loss.html}} - an L$_{1}$ loss that is smooth if $|x-y|<\beta$, where $\beta=1\sigma$ and $\sigma$ is the standard deviation. The loss for $|x-y|<\beta$ was further increased by a factor of $10$ to aid the efficiency of the training convergence of the model. It can be viewed as a combination of L$_{1}$ and L$_{2}$ losses (behaves as L$_{1}$ if the absolute value is high or as L$_{2}$ if the absolute value is low).
\end{itemize}

Besides the network itself, the reasons behind the choices of Optimizer, Activation Function, and the Loss Function were the broad applicability, availability, and success of these techniques. In addition, we performed cross-validation over available PyTorch functions to solidify our choices. These parameters were sufficient for off-diagonal terms of $\tau_{ij}$ of 3D MHD turbulence. However, physical conditions had to be enforced in order to model diagonal terms and ultimately predict $P_{turb}$.

\subsubsection{Diagonal Terms (PIML)}
\label{sec:piml}

If calculated directly, Reynolds stress is defined by:
\begin{gather}
    \tau_{ij} = \widetilde{u'_iu'_j} = \widetilde{u_iu_j}-\tilde{u}_i\tilde{u}_j
    \label{eq:raw_tn}
\end{gather}
\noindent where $u'_i$ is a velocity fluctuation component, and $\tilde{x}$ is the spatial average. Thus, for diagonal terms a realizibility condition is defined as $\tau_{ii} > 0$ \citep{schumann1977}, making distribution of diagonal tensor components asymmetric. While the model in Section \ref{sec:3dcnn} excelled at quasi-symmetric distribution prediction, it struggled with asymmetric distributions. Further analysis of this will be covered in Section \ref{sec:dynamic_results}.

\begin{itemize}
    \item \textbf{Model}: 3D CNN encoder as described in Section \ref{sec:3dcnn} with physics-informed (PI) layers. The encoder implicitly predicts velocity fluctuations ($u_i'$), then PI layers calculate $u_i'^2$ to enforce $\tau_{ii} > 0$ and filter to find the mean as per Eq. \ref{eq:raw_tn}.
    \item \textbf{Optimizer}: Adam Optimizer - same as in Section \ref{sec:3dcnn}.
    \item \textbf{Activation function}: LogSigmoid (stationary) and Tanhshrink\footnote{\href{https://pytorch.org/docs/stable/generated/torch.nn.Tanhshrink.html}{pytorch.org/docs/stable/generated/torch.nn.Tanhshrink.html}} (dynamic) - the latter showed a faster model convergence rate for the dynamic turbulence. The function is defined by \[f(x) = x-tanh(x)\]
    \item \textbf{Loss function}: Custom - a dynamic combination of SmoothL1Loss (point-to-point) and Kolmogorov-Smirnov (KS) Statistic \citep{ks1951} losses.
\end{itemize}

The combined loss function was designed to compound the advantages of the point-to-point and statistical losses. Further discussion can be found in the Results Section \ref{sec:dynamic_results}.

The {\sf PyTorch} implementation of this PIML model used for diagonal terms along with the 3D CCSN data sampled down to $17^3$ is provided as part of the {\sf Sapsan} package\footnote{\href{https://github.com/pikarpov-LANL/Sapsan/wiki/Estimators\#physics-informed-cnn-for-turbulence-modeling}{github.com/pikarpov-LANL/Sapsan/wiki/Estimators\#physics-informed-cnn-for-turbulence-modeling}}.

\subsection{Datasets}
\label{sec:datasets}
In machine learning, the predictions will only be as good as the training data. With the goal of testing our algorithm on a broader range of physical conditions, we diversified by using stationary and dynamic turbulence datasets. A quick parameter overview of both datasets can be found in Table \ref{tab:data_pars}.

\begin{itemize}
    \item \textbf{Stationary}: high resolution statistically stationary, isotropic, forced, incompressible MHD turbulence dataset\footnote{\href{http://turbulence.pha.jhu.edu/Forced_MHD_turbulence.aspx}{turbulence.pha.jhu.edu/Forced\_MHD\_turbulence.aspx}} from Johns Hopkins Turbulence Database (JHTDB) \citep{jhtdb2008}. It has a low Reynolds number fluctuating around $Re\sim186$.
    
    \item \textbf{Dynamic}: evolving highly-magnetized CCSN turbulence dataset by \cite{moesta2015}. It is a 3D DNS of an MHD CCSN. The dataset has been developed to study the effects of magneto-rotational instability (MRI) in growing turbulence in a CCSN scenario to aid the explosion, aiming to prove the plausibility of CCSN to be progenitors of LGRBs and magnetars \citep{moesta2015}. The needed resolution is high, so it is not a global simulation, only tracking the first $\sim10\:ms$ after the core bounce to see the development of turbulence. In addition, only a quarter of the star close to the core has been simulated ($66.5\:km$ in $x$ and $y$ and $133\:km$ in $z$), maintaining a $90^{\circ}$ rotational symmetry in the $xy$ plane. To fit our memory constraints, we used the dataset with $\Delta x=200\:m$ resolution that exhibits mild turbulence.
\end{itemize}

\begin{table}[]
    \centering
    \begin{tabular}{c|c|c}
        & Stationary & Dynamic \\
        \hline
        Resolution & $1024^3$ & $347^3$\\
        $t_{tot}$ & 2.56 & 10.3 [ms] \\
        $\delta t$ & $\num{2.5e-4}$ & $\num{5e-4}$ [ms] \\
        $\Delta t$ & $\num{2.5e-3}$ & $\num{2.4e-2}$ [ms] \\
        $k_{max}$ & 482 & 348 \\
        $\overline{KE/E_{tot}}$ & $\sim0.5$ & $\sim0.9$\\
        $\overline{E_B/E_{tot}}$ & $\sim0.5$ & $\sim0.1$\\
        \hline
        $\sigma$ & 16 & 9\\ 
    \end{tabular}
    \caption{Parameters of the statistically \textit{stationary} (JHTDB) and \textit{dynamic} CCSN \citep{moesta2015} turbulence datasets. The time values for \textit{stationary} dataset is in normalized numerical units amounting to 1 large-eddy turnover time. $t_{tot}$ is the total simulation time, $\delta t$ is a timestep, $\Delta t$ is checkpoint time separation, $k_{max}$ is the maximum Fourier mode, $\overline{KE/E_{tot}}$ and $\overline{E_B/E_{tot}}$ are the time-averaged fractions of kinetic and magnetic energy with respect to the total energy. $\sigma$ is the Gaussian filter standard deviation we applied during data preparation. Lastly, the spatial resolution of the \textit{Dynamic} dataset is $\Delta x=200\:m$.}
    \label{tab:data_pars}
\end{table}

\subsubsection{Data Preparation}
\label{sec:data_prep}

\begin{figure*}
    \centering
    \includegraphics[width=\linewidth]{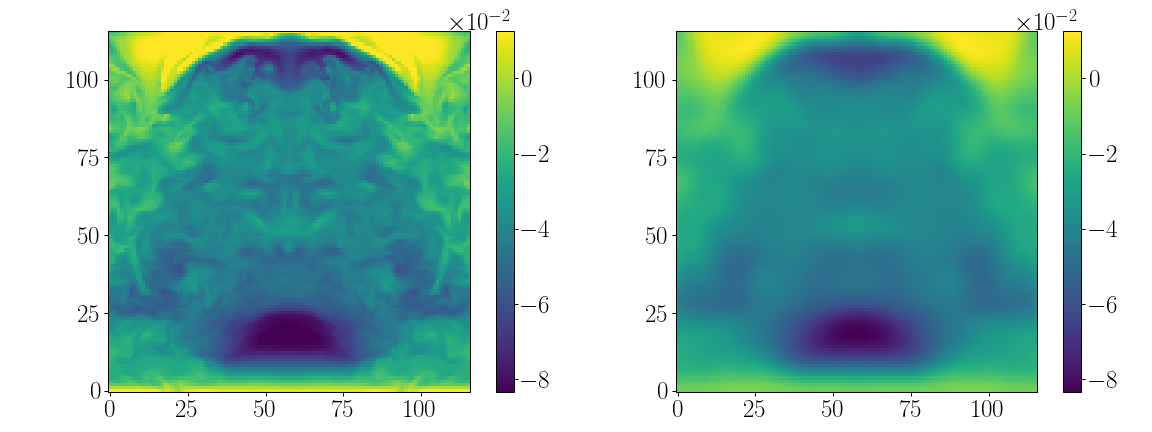}
    \caption{Box filter with a Gaussian kernel ($\sigma=9$) was used to filter the data. Above is a slice of sampled $u_x$, down to (116,116,116); filter was applied prior to sampling. \textbf{Left:} Original, \textbf{Right:} Filtered}.
    \label{fig:gaussian_filter}
\end{figure*}

In order to reflect a realistic low-fidelity environment of the CCSN simulations, we chose to apply Gaussian filter \citep{carati2001} as described in Section \ref{sec:filtering}. Standard deviation $\sigma$ of the filter (Eq.\ref{eq:gauss}) for each dataset was chosen to provide similar levels of filtering for both \textit{stationary} and \textit{dynamic} datasets, with exact $\sigma$ specified in Table \ref{tab:data_pars}. An example of the filtering used is shown in Fig. \ref{fig:gaussian_filter}. For both datasets, derivatives were calculated, and the filter was applied at the highest resolution available. Then, the data was equidistantly sampled down to $116^3$ to fit the hardware memory constraints. The exact data preparation procedures are summarized below:
\begin{enumerate}
    \item Calculate derivatives of $[u, B]$
    \item Calculate Reynolds stress tensor $\tau$ using Eq.\ref{eq:velocity_stress}
    \item Apply a \textit{Gaussian filter} to the original data to get $[\tilde{u}, \widetilde{du}, \tilde{B}, \widetilde{dB}]$
    \item Sample the data equidistantly down to 116$^3$
    \item Use the sampled quantities of $[\tilde{u}, \widetilde{du}, \tilde{B}, \widetilde{dB}]$ as the model \textit{input}
    \item Use each $\tau_{ij}$ component as the model \textit{output}
\end{enumerate}

\section{Results \& Discussion}
\label{sec:results}
\subsection{Stationary Turbulence}

\begin{figure*}[!h]
    \centering
    \includegraphics[width=\linewidth]{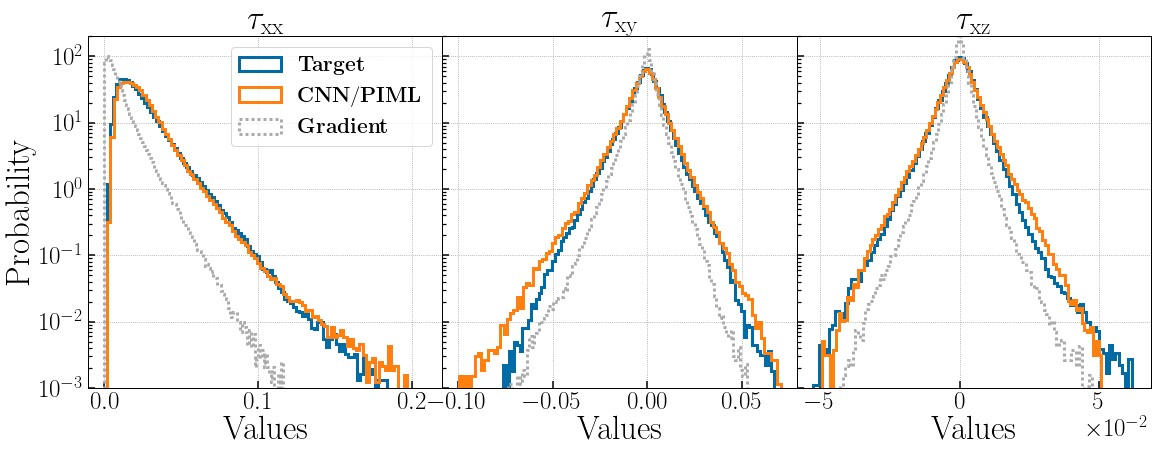}
    \caption{Prediction of $x$ components of $\tau$ at $t=1000$ of JHTDB MHD dataset in normalized numerical units. \textit{Blue} is the original target data, \textit{Orange} is prediction of the CNN or PIML models for off-diagonal and diagonal terms respectively, and \textit{Gray} is the result of the gradient subgrid model as per Section \ref{sec:gradient}.}
    \label{fig:jhtdb}
\end{figure*}

\begin{figure*}[ht]
    \centering
    \includegraphics[width=\linewidth]{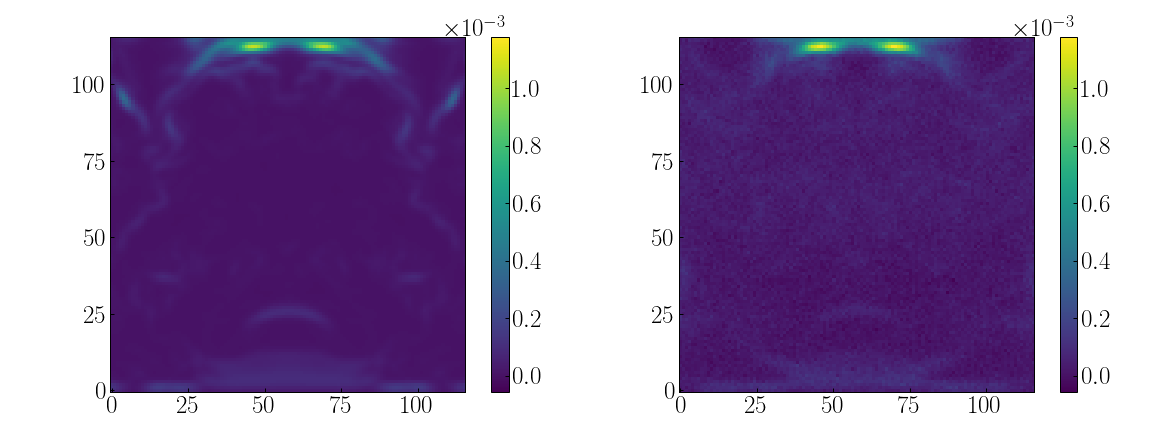}
    \caption{Slices presenting spatial distribution of 3D stress tensor component at $t=7.55 \: ms$, sampled down to $116^3$. \textbf{Left:} Target $\tau_{xx}$, \textbf{Right:} PIML Prediction of $\tau_{xx}$. Statistical distribution of the above can be found in Fig. \ref{fig:combined} ($1^{st}$ row, $2^{nd}$ column)}
    \label{fig:it501}
\end{figure*}

Though the dataset is evolving spatially, statistics remain stationary in the JHTDB MHD dataset we used. Our CNN and PIML models for off-diagonal and diagonal terms respectively outperform the traditional gradient subgrid model. Fig.\ref{fig:jhtdb} presents predictions of the $x$ components of the Reynolds stress: [$\tau_{xx},\tau_{xy},\tau_{xz}$] at $\Delta t = 1000$, which is near the end of the simulation. Prediction performance of the $y$ and $z$ components remained comparable to the $x$ components; hence those plots were omitted.

For training, we used the first 4 checkpoints ($\Delta t$), separated by 10 time-steps ($\delta t$) each. Exact data preparation was performed as per Section \ref{sec:data_prep}. Our PIML method especially excelled at predicting $\tau_{xx}$ that is consequently important for $P_{turb}$ calculation, while the gradient model completely misses the peak and the overall turbulent distribution. Next, $\tau_{xy}$ matches the bulk of the data but overpredicts the outliers. Note that $y-axis$ is on a log scale; hence the actual error remains minimal. As for $\tau_{xz}$, the CNN model does not show any particular weaknesses.

The point of this exercise was to test the reliability of the CNN algorithm when applied to a changing spatial distribution. Since CNNs parse a kernel through the data-cube, it is not trivial to assume statistical consistency in the predictions based on the evolving spatial distribution. Nonetheless, the statistically stationary dataset did not require significant tuning to achieve its results and served more as a consistency check of our algorithms before moving to a dynamic dataset.

\begin{figure*}[hbtp]
    \centering
    \includegraphics[width=\linewidth]{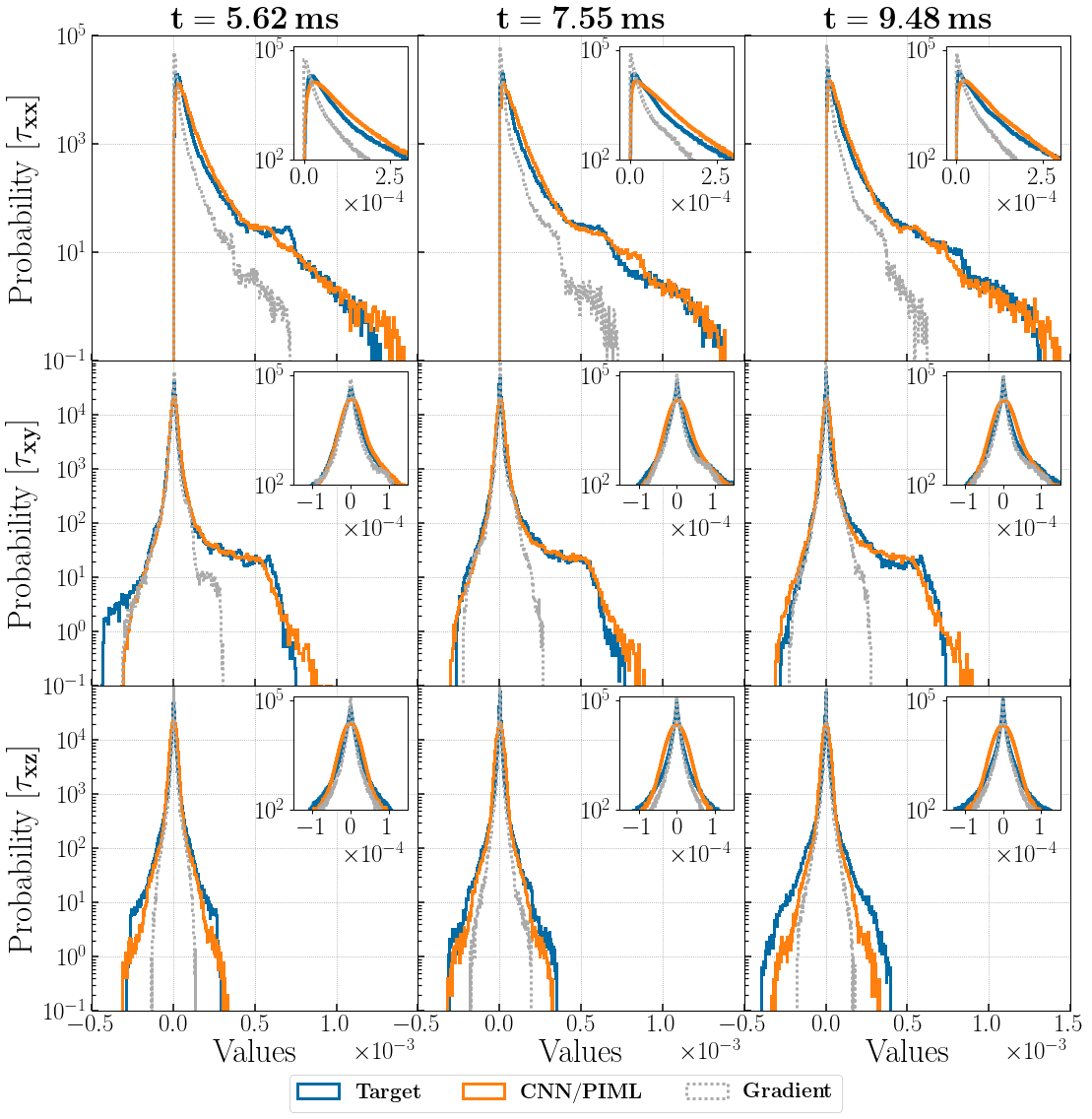}
    \caption{Statistical distributions of the stress tensor component, where the values are in units of $u^2$. \textbf{Rows}: $[\tau_{xx}, \tau_{xy}, \tau_{xz}]$ components; \textbf{Columns}: simulation time $[5.62, 7.55, 9.48] ;\ ms$. \textbf{\textcolor[RGB]{0, 107, 164}{Blue}}: Target $\tau$ distribution; \textbf{\textcolor[RGB]{255, 90, 0}{Orange}}: CNN prediction; \textbf{\textcolor[RGB]{80,80,80}{Gray Dotted}}: gradient turbulence subgrid model of the form $\tau_{ij}=1/12 \Delta^2\delta u_{ik}\delta u_{jk}$ using Einstein notation, where $\Delta$ is filter width and $u$ is velocity.}
    \label{fig:combined}
\end{figure*}

\subsection{Dynamic Turbulence}
\label{sec:dynamic_results}

The ulimate goal of our study was to test the models on a more realistic astrophysical dataset. While DNS CCSN simulation from \citep{moesta2015} has its limitations, it is the best-resolved turbulence dataset investigating CCSN. Figure \ref{fig:combined} presents predictions of $\tau_{x}$ components in the second half of the simulation, $t=[5.62,\;7.55,\;9.48]\:ms$. The results remain consistent with statistically stationary JHTDB predictions. The gradient model continuously underpredicts the Reynolds stress distribution, performing especially poorly at capturing the outliers. Predictions of $\tau_{y}$ and $\tau_{z}$ components are comparable in accuracy across all timesteps, hence were omitted from the plot. Figure \ref{fig:it501} presents an example of spatial distribution prediction, i.e. slice of the datacube at $t=7.55\:ms$.

The key to capturing the dynamics was to train across a wide range of checkpoints covering the first half of the simulation, $t<5\:ms$. We were able to optimize the results using 9 equally distant checkpoints; the evolution of normalized PDFs of $\tau_{xy}$ from the exact checkpoints used in training can be seen in Figure \ref{fig:training}. A diversified training dataset as such helped prevent overfitting while maintaining the flexibility of the model to predict the future timesteps ($t>5\:ms$). 

Thus far, the CNN methods worked well for off-diagonal terms, and our PIML enforced realizability condition ($\tau_{ii}>0$) for the diagonal terms. However, diagonal terms of the dynamic turbulence dataset presented another challenge - asymmetric statistical distributions. CNN with a point-to-point loss such as SmoothL1Loss has shown its robust performance at predicting quasi-symmetric distributions. This includes predictions of diagonal terms in JHTDB stationary data ($\tau_{xx}$ in Figure \ref{fig:jhtdb}), where due to the shift of the peak, the distribution can be classified as quasi-symmetric. However, in the dynamic dataset, the peak is near the origin, making the distribution acutely asymmetric. As a result, while accurately capturing the outliers, the previous approach failed to predict the correct peak position, i.e., the bulk of the data. To remedy this behavior, we developed a custom loss function combining a point-to-point SmoothL1Loss with a loss based on the Kolmogorov-Smirnov statistic ($KS_{stat}$). The latter metric is the maximum distance between the two cumulative distribution functions (CDFs), i.e., how far apart the two distributions are from one another.

SmoothL1Loss excelled at predicting distribution outliers while struggling to determine the peak position. On the other hand, KS loss excelled at predicting the bulk of the data, including the peak position, by minimizing the distance between the input and target distributions but struggled with the outliers. As a result, the two losses were combined in a dynamic fashion. The model first minimized SmoothL1Loss to get the overall distribution shape, particularly the outlier portion, and then minimized KS loss to shift the peak into the right position. The results can be seen in the top row of Figure \ref{fig:combined}, with the detailed peaks presented in a separate box within each plot. Further details on the training loss behavior are provided in Appendix \ref{appendix:loss}. While we primarily stressed accurate prediction of the statistical distribution, another benefit of not using an exclusively statistical loss is a sound spatial prediction, as shown in Figure \ref{fig:it501}.

\begin{figure*}
    \centering
    \includegraphics[width=\linewidth]{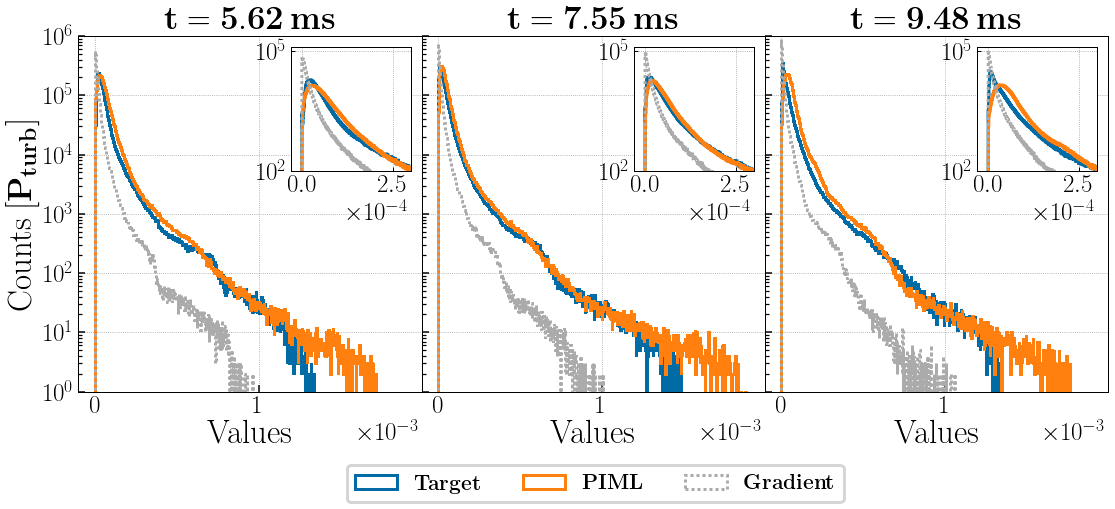}
    \caption{The plot presents an unnormalized distribution of $P_{turb}$ as time evolves. $P_{turb}$ is in units of $u^2$. We compare the performance of our PIML model with the gradient turbulence subgrid model.}
    \label{fig:piml_pturb_evol}
\end{figure*}

The leading deliverable of $\tau_{ij}$ predictions is the ability to calculate turbulence pressure via Eq.\ref{eq:pturb}. Thus, any deviation in the peak of $\tau_{ii}$ is further exacerbated when computing $P_{turb}$. As an example of our PIML model performance, we present $P_{turb}$ prediction calculation at $t=[5.62, 7.55, 9.48]\:ms$, as shown in Figure \ref{fig:piml_pturb_evol}. There, the trace was taken of the \textit{sorted} $\tau_{ii}$ components. During this operation, the \textit{spatial} distribution of the $P_{turb}$ is lost, though it is not required for our main goal: accurate prediction of the total pressure due to turbulence in the region and its \textit{statistical} distribution. This is due to the convection region being extremely under-resolved, while it is responsible for supplying $P_{turb}$ to the stalled shock for the potential explosion in the global CCSN simulations. Thus, the astrophysical question is reduced to a binary one: will the stalled shock move outwards (explosion) or inwards (black hole). Consequently, the accuracy of the total $P_{turb}$ in the convection region becomes the most significant while alleviating the need for accurate prediction of the spatial distribution.

\begin{figure}
    \centering
    \vspace{0.3cm}
    \includegraphics[width=\linewidth]{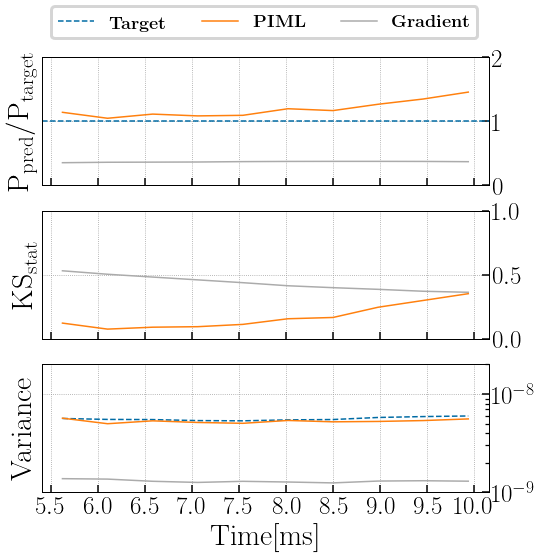}
    \caption{Performance metrics of the PIML vs. gradient subgrid model. The \textit{Top} panel is the ratio of the total turbulent pressure calculated from the model prediction to the target dynamic 3D DNS CCSN data, \textit{Middle} is the Kolmogorov-Smirnov statistic, and \textit{Bottom} panel is variance. In total, metrics at 10 checkpoints equally separated in time are presented in the plots.}
    \vspace{0.2cm}
    \label{fig:pturb_stats}
\end{figure}

The performance of the PIML method shows significant advantages over the gradient model predicting the over distribution, the outliers, and the peak position. Its performance does deteriorate with time, as can be seen by the slight peak shift in Figure \ref{fig:piml_pturb_evol}, right plot at $t=9.48\:ms$. Quantitatively, performance metrics to compare PIML and the gradient model predictions are summarized in Figure \ref{fig:pturb_stats}. The \textit{Top} panel shows that the total $P_{turb}$ calculated from the PIML predictions over-predicts the target ground truth (the dynamic 3D DNS CCSN data) by $\sim5\% - 35\%$ depending on the future prediction time, resulting in a $\sim 19\%$ deviation on average. This is a significant advantage over $\sim63\%$ under-prediction error of the gradient model that will fail to supply sufficient $P_{turb}$ to re-energize the stalled shock to explode the star. This means that by using the PIML method, $P_{turb}$ could reach on average $\sim36\%$ of the gas pressure instead of the estimated $\sim30\%$  in \cite{nagakura19}, making it easier for the star to explode.


The \textit{Middle} panel of Figure \ref{fig:pturb_stats} shows $KS_{stat}$ for the PIML method to degrade to gradient model level at the far-future checkpoints. This large discrepancy in the Target and PIML CDFs is due to the slight peak shift of the prior PIML prediction. While $KS_{stat}$ is an important metric used in our custom loss function, it does not disqualify PIML's advantages over the conventional gradient turbulence model.

Lastly, the \textit{Bottom} panel presents consistent variance between the Target and PIML results. The predicted distribution stays consistent in its dispersion, i.e., bulk and outlier distribution, which cannot be said about the gradient model results. Thus the PIML approach has an advantage in modeling the small-scale eddies that, in turn, can grow into large scales to provide the dominant fraction of the $P_{turb}$ to re-energize the stalled shock as the simulation evolves.

\section{Conclusion}
\label{sec:conclusion}

The study of CCSN requires an accurate treatment of turbulence, and yet conventional subgrid turbulence approaches are unreliable. A DNS treatment of turbulence in global 3D CCSN simulations is not achievable with the current computational resources, thus the calculations are typically done via ILES. Although they can capture the effects of large-scale flows with relative accuracy, these simulations neglect the turbulent pressure ($P_{turb}$) entirely, relying on numerical artifacts to represent its effect. Building upon prominent ML techniques used in the industry, we have developed PIML networks to increase the predictive accuracy of Reynolds stress ($\tau_{ij}$), the trace of which is $P_{turb}$. $P_{turb}$ is thus the main deliverable of this paper that can be used in a CCSN simulation in aid of re-energizing the stalled shock and exploding the star. 

Our PIML approach consistently outperformed a conventional gradient subgrid model for both \textit{stationary} and \textit{dynamic} turbulence datasets. It resulted in a $\sim19\%$ PIML average error of the total $P_{turb}$ in comparison to $\sim63\%$ of the gradient model. In addition, our method has excelled at predicting the outliers of both $\tau_{ij}$ and $P_{turb}$, which are important for dynamic simulations to investigate the turbulent growth. Given the flexibility of ML algorithms used, these results should be reproducible across HD and MHD CCSN simulations, which we are currently investigating for the next publication. That being said, the performance of the ML models deteriorates further in time predictions are made, which is to be expected with a CNN-based approach. While at its worst, it continued to take the lead over the gradient model, temporal and overall performance can be further improved in our future work with the inclusion of recurrent neural network (RNN) layers in the models or by utilizing physics-informed neural operators (PINO) \citep{li2021, rosofsky2022}.

Furthermore, a broader application of the ML model can suffer from the data-model inconsistency when integrating a trained model within CCSN codes. The distribution discrepancies between the training dataset and the newly simulated grids, as well as the numerical errors, can lead to an exponential error growth in the predictions \citep{beck21}. Regularization of the model prediction can improve its stability. Our future work will be investigating the approaches to tackle this potential issue.

This paper has been our first attempt at studying the generalizability of ML methods for studying turbulence over different physical regimes. In the future, we would like to delve deeper into this topic, employing other 3D MHD CCSN datasets. Specifically, here we used a DNS MHD CCSN dataset of a single star, while we would like to expand the study to both HD and MHD models over a wide range of CCSN progenitor masses (from $8\:M_{\odot}$ to $25\:M_{\odot}$) that exhibit great variation in their physical engines. In the next paper, we will present our current implementation of the evolving turbulent pressure term trained on 3D simulation data into 1D CCSN models to study a large parameter space of progenitors to understand its impact on the CCSN explosion rates.

\section{Acknowledgment}
The research presented in this paper was supported by the Laboratory Directed Research and Development program of Los Alamos National Laboratory (LANL), a LANL Center of Space and Earth Science student fellowship, and the DOE ASCR SciDAC program. This research used resources provided by the LANL Institutional Computing Program, which is supported by the U.S. Department of Energy National Nuclear Security Administration under Contract No. 89233218CNA000001. We would also like to thank Philipp M{\"o}sta for providing the 3D MHD CCSN dataset used throughout this paper, and Jonah Miller for insightful discussions that helped develop the ML methods.

\bibliographystyle{aasjournal}
\bibliography{bibliography}

\begin{appendices}
\label{sec:appendix}

\section{Training Features}
\label{appendix:loss}

\begin{figure*}[!h]
    \centering
    \includegraphics[width=\linewidth]{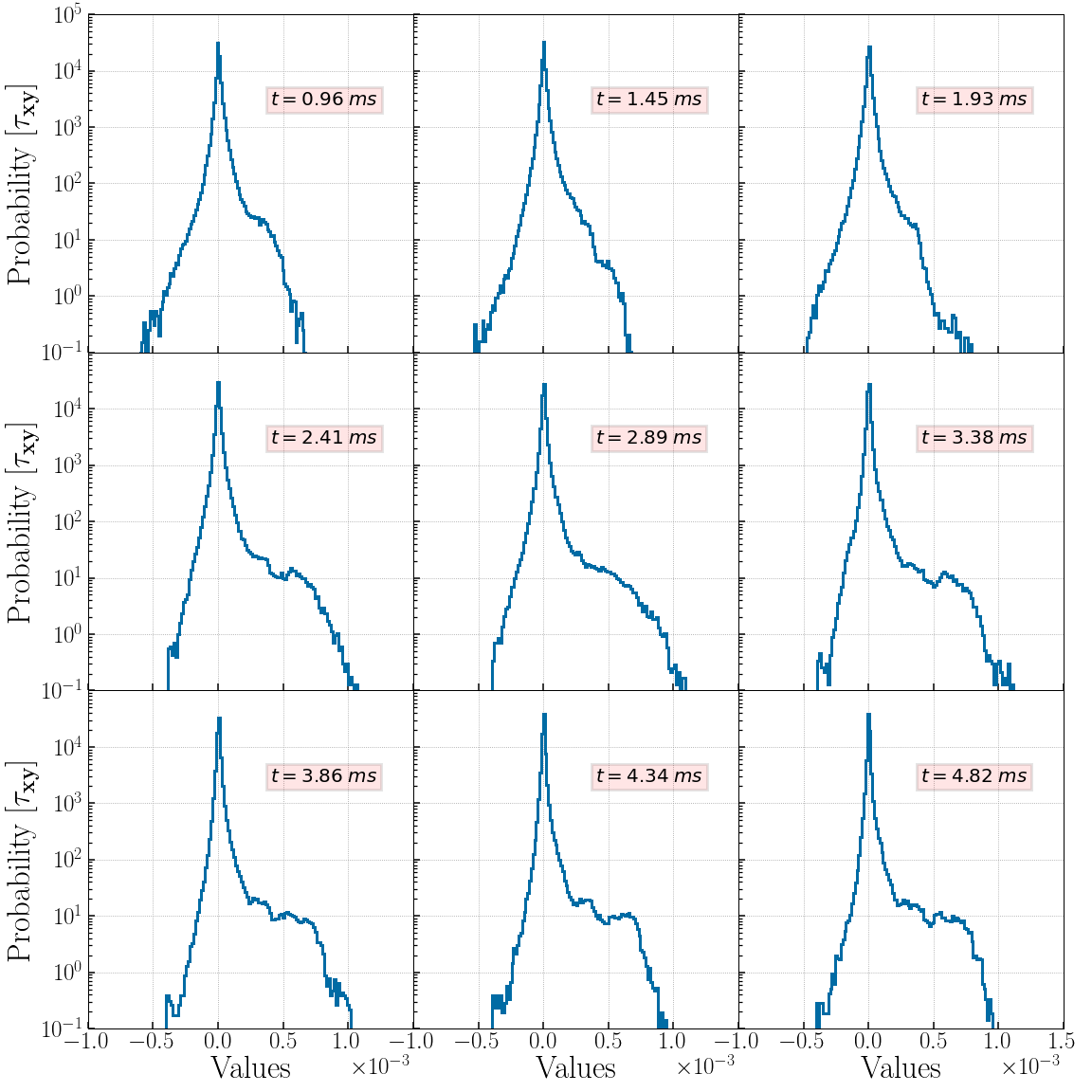}
    \caption{Evolution of the stress tensor component's ($\tau_{xy}$) statistical distribution in the first half of the simulation ($t < 5\:ms$). These exact checkpoints, 9 in total, where used as a \textit{target} to train the CNN network.}
    \label{fig:training}
\end{figure*}

The ML models were trained on $u,\;du,\;B,\;dB$ as the input features and $\tau_{ij}$ as the target feature: a model per tensor component. Figure \ref{fig:training} presents an example of how $\tau_{xy}$ evolves at $t<0.5\:ms$, following exact checkpoints used for training. The other $\tau_{ij}$ components follow a similar level of dynamics. This provided a sufficient level of diversity in the training dataset to prevent model overfitting and aid flexibility of the model.

\section{Training Loss}
\label{appendix:loss}

We developed a custom loss function ($l$) that combines a point-to-point (SmoothL1Loss, i.e. $L1$) with a statistical loss (Kolmogorov-Smirnov statistic, i.e., $KS_{stat}$) in a dynamic fashion. The goal was to force the model to minimize $L1$ loss first to get the overall distribution shape. Then, the model prioritizes minimizing $KS_{stat}$ to shift the peak into its correct position. 

\begin{figure*}[h]
    \centering
    \includegraphics[width=0.7\linewidth]{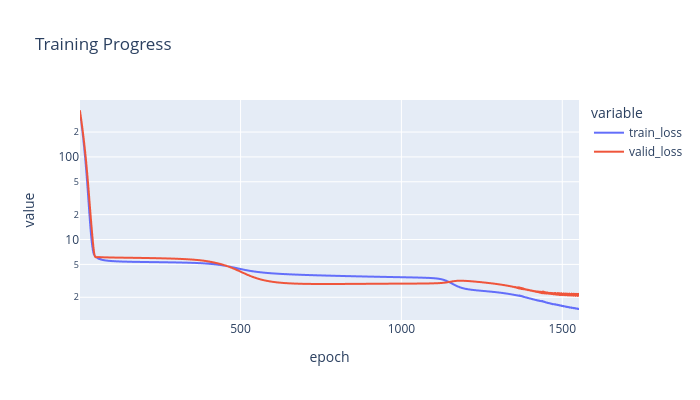}
    \includegraphics[width=0.7\linewidth]{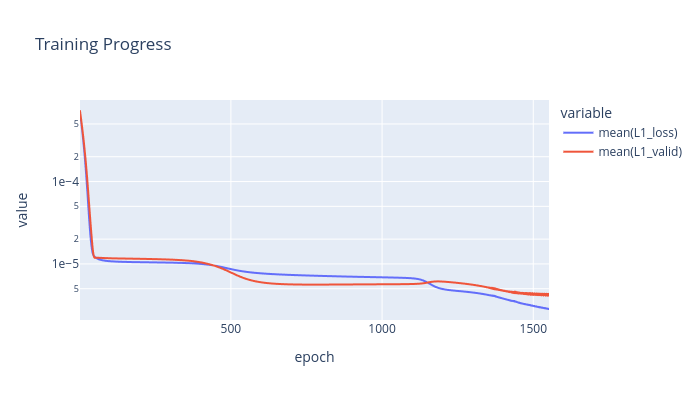}
    \includegraphics[width=0.7\linewidth]{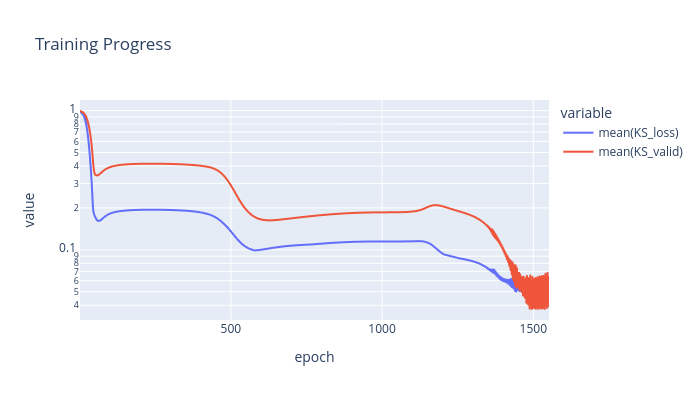}
    \caption{Training loss evolution: \textit{Top} is the actual loss of the model that consists of a combined L$_{1}$ and KS loss components, \textit{Middle} is the L$_{1}$ loss component, \textit{Bottom} is the $\mathrm{KS_{stat}}$ loss component.}
    \label{fig:loss}
\end{figure*}

The two losses operate on different scales during training: $L1$ loss can span a range of $\sim10^4$ (from $10^{-4}$ down to $10^{-8}$), while $KS_{stat}$ ranges $\sim10^1$ (from $10^0$ down to $10^{-1}$) before overfitting. To account for such differences, we first normalized the losses and then applied a scaling factor $\alpha$ to prioritize $L1$ until the general PDF shape had been learned. Given our training data, at $L_1 \; loss < 10^{-6}$ this condition was satisfied, making $\alpha = l_{init}/10^{-6}$. Thus, $L1$ loss is heavily prioritized for the first several orders of magnitude, decreasing the combined loss ($l$), then sharing an equal weight with $KS_{stat}$. This results in $l$ to follow $L1$ loss's training dynamic as shown by Figure \ref{fig:loss} $Top$ and $Middle$ plots. In summary, the two losses are combined as follows:
\begin{gather}
    l = 0.5\left(\alpha\frac{\widetilde{L1}}{L1_{init}}\right)+0.5\frac{\widetilde{KS_{stat}}}{KS_{stat,init}}
\end{gather}
\noindent where $\widetilde{X}$ is a spatial average of $X$, and $\widetilde{X}/X_{init}$ is effectively a normalized quantity, since $X_{init}\approx X_{max}$ for losses. 

Once $KS_{stat}$ becomes important, the peak is being shifted, and we introduce an early stopping condition to prevent overfitting. Figure \ref{fig:loss} $Bottom$ plot presents the training evolution of the $KS_{stat}$ loss component of $l$. After train loss ($blue$) and validation loss ($red$) cross at $\sim\num{4e-2}$, train loss decreases exponentially while validation loss increases exponentially. This indicates that the model is overfitting. Thus, the early stopping condition was set to $\sim\num{4e-2}$, based on the crossing value of train and validation losses.

More details on the application and reasoning behind the combined loss can be found in Section \ref{sec:dynamic_results}.

\section{CNN Encoder}
\label{appendix:cnn}

\begin{figure*}[h]
    \centering
    \includegraphics[width=0.6\linewidth]{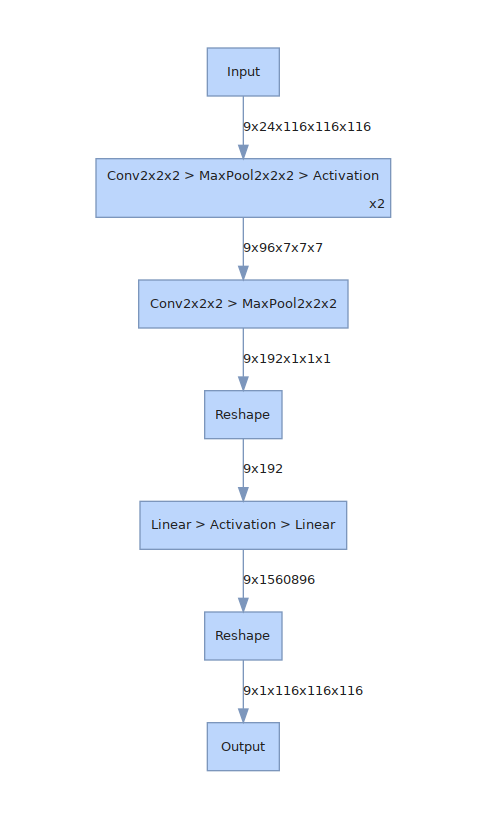}
    \caption{Graph of the CNN encoder used in all models for tensor component prediction. The activation function is either LogSigmoid or ShrinkTanh depending on the tensor component type. Further information can be found in Section \ref{sec:3dcnn}}
    \label{fig:3dcnn_graph}
\end{figure*}

We present a graph of the CNN encoder we used in Figure \ref{fig:3dcnn_graph}. Data shape is noted at each arrow, akin to what was used to produce our results throughout the paper. For input and output, the shape is formatted as $[N,C,D,H,W]$ where $N$ is the number of batches, $C$ represents channels, i.e., features, and [$D$,$H$,$W$] stand for depth, height, and width of the data. The notation is in agreement with {\sf PyTorch} documentation. The graph was produced with {\sf Sapsan}\footnote{\href{https://github.com/pikarpov-LANL/Sapsan/wiki/Model-Graph}{github.com/pikarpov-LANL/Sapsan/wiki/Model-Graph}}.

\end{appendices}
\end{document}